\documentclass[reprint,aps,prx,longbibliography,superscriptaddress]{revtex4-2}

\usepackage{graphicx}
\usepackage[utf8]{inputenc}
\usepackage{xcolor}
 \definecolor{goodgreen}{rgb}{0.1,0.5,0}
 \definecolor{goodred}{rgb}{0.7,0,0}
\usepackage[colorlinks,urlcolor=blue,citecolor=goodgreen,linkcolor=goodred]{hyperref}
\usepackage{siunitx}
\usepackage{upgreek}
\usepackage{newtxtext,newtxmath}

\renewcommand{\text}[1]{\ensuremath{\mathrm{#1}}}
\newcommand{\un}[1]{\ensuremath{\,\textrm{#1}}}
\newcommand{\Vg}{\ensuremath{V_\text{g}}}

\newcommand{\vg}{\Vg}

\newcommand{\vsd}{\ensuremath{V_{\text{sd}}}}
\newcommand{\moss}{MoS\textsubscript{2}}
\newcommand{\sioo}{SiO\textsubscript{2}}

\DeclareSIUnit\bar{bar}

\begin{document} 

\title{Reproducible Ohmic bismuth contacts to \moss\ nanotubes and nanoribbons}

\author{Robin T. K. Schock}
\author{Stefan Obloh}
\author{Korbinian Fink}
\author{Matthias Kronseder}
\affiliation{Institute for Experimental and Applied Physics,
University of Regensburg, 93040 Regensburg, Germany}

\author{Matja\v{z} Malok}
\author{Maja Rem\v{s}kar}
\affiliation{Solid State Physics Department, Institute Jo\v{z}ef Stefan, 1000
Ljubljana, Slovenia}

\author{Andreas K. H\"uttel}
\email[]{mail@akhuettel.de}
\affiliation{Institute for Experimental and Applied Physics,
University of Regensburg, 93040 Regensburg, Germany}

\begin{abstract}
Attaching metallic contacts to transition metal dichalcogenide nanostructures
and in particular to \moss\ has posed significant challenges over the past
years. For \moss\ nanotubes and nanoribbons, a highly promising material for
field effect transistors as well as quantum electronic devices, this is even
more the case due to the small, curved surface. So far all attempts there have 
led to a wide scatter of contact resistances on the same chip. Recently, for 
quasi two-dimensional, flat \moss\ flakes, the use of semimetals has led to a 
breakthrough, making transparent and Ohmic contacts possible. Here, we 
demonstrate the steps required to reproducibly fabricate contacts to single, 
vapor phase grown \moss\ nanotubes and nanowires. All devices display finite
room-temperature two-point resistances in absence of gating, with a median 
value of \qty{340}{\kohm} in a large fabrication series. A detailed analysis 
elucidates the impact of the different fabrication changes.
\end{abstract}

\maketitle 

The isolation of few- and monolayer 2d materials such as graphene 
\cite{novoselov_electric_2004} or transition metal dichalcogenides 
\cite{novoselov_two-dimensional_2005} has led to an avalanche of research 
regarding both classical semiconductor applications \cite{das_transistors_2021, 
dorow_exploring_2023, chou_status_2023} and quantum electronics 
\cite{jing_gate-controlled_2022, liu_2d_2019, qiu_recent_2021}. Nevertheless, 
both commercial chip fabrication where integrated transistor channels become 
narrower and narrower \cite{hills_modern_2019, si_carbon-nanotube-based_2024} 
and quantum electronics with the manyfold focus on one-dimensional physics 
\cite{science-stern-2013, science-frolov-2026} also point out the intrinsic 
geometric advantage of quasi-1d nanostructures such as nanotubes, nanoribbons, 
or nanowires. This has driven recent fundamental research on complex 1d 
nanomaterials beyond carbon, including even so-called coaxial heterostructures 
\cite{xiang_one-dimensional_2020}.

Molybdenum disulfide \moss, mostly known as 2d material and semiconductor
\cite{novoselov_two-dimensional_2005}, also allows the growth of long,
straight, low-defect density nanotubes and nanoribbons via chemical vapor
deposition \cite{remskar1996}. These 1d nanomaterials of large aspect ratio
display outstanding optical properties \cite{kazanov_multiwall_2018,
remskar2022} and have shown clear promise in first field effect transistor
devices \cite{fathipour2015}. The central challenge for electronic measurements
is however the reliable fabrication of Ohmic contacts. The formation of
Schottky barriers at the metal-semiconductor interface
\cite{zphys-schottky-1939, tung_physics_2014}, often caused or enhanced by Fermi 
level pinning \cite{louie_electronic_1976, monch_valence-band_1998,
tung_physics_2014}, has required significant efforts to overcome 
\cite{nl-gong-2014, kim_fermi_2017, jpcc-sotthewes-2019}.

For the 2d morphology of \moss, i.e., multi- and monolayers, recently
breakthrough results have been achieved using semimetal contact layers, in
particular bismuth \cite{shen2021a, schock2023} and antimony \cite{li2023}. A
transfer of this technique to 1d morphologies was partially successful
\cite{schock2023, schock2025}, with bismuth contacts leading to transparent 
devices, but at relatively low yield and still a wide scatter of device 
resistance values.

Here, we demonstrate the reproducible fabrication of \moss\ nanotube and
nanoribbon devices, by combining our experience from previous work 
\cite{schock2023, schock2025} with several fabrication techniques developed in
other material systems. The results are characterized using room temperature
resistance measurements and demonstrate a reduction of the 2-point resistance 
$R_2$ scatter by several orders of magnitude, with all measured nanotube and 
nanoribbon segments displaying $R_2 < \qty{30}{\Mohm}$. Given the extensive 
interest world-wide in \moss\ integrated circuits, and the chance provided here 
to reduce device dimensionality and with it the potential area footprint of 
transistors, this development is not just interesting for quantum devices but 
also of clear technological impact as well.

\section*{Raw material and device fabrication}
\begin{figure*}[tbp]
\begin{center}
\includegraphics{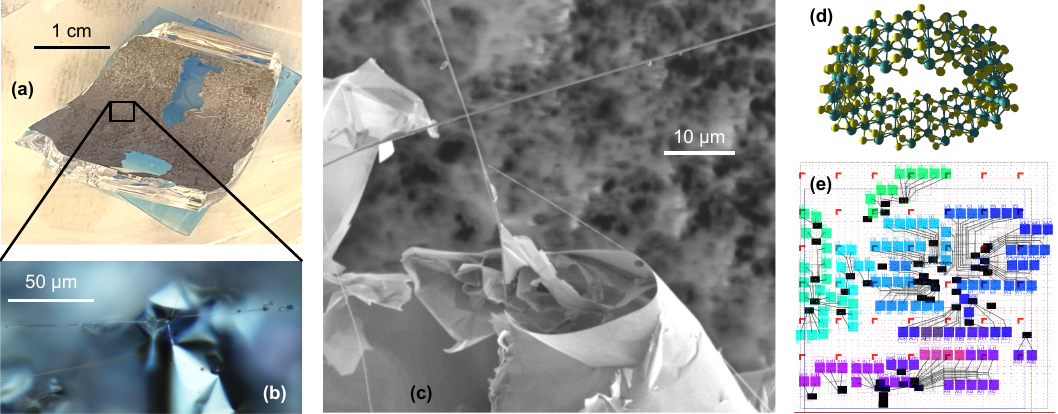}
\end{center}
\caption{{\bf Raw material and device geometry.
\quad}
{\bf a,} As-grown \moss\ nanomaterial deposited on the wall of a broken growth 
quartz glass ampoule.
{\bf b,} Optical microscope image of the \moss\ material, with a one-dimensional
nanostructure (nanotube or nanoribbon) surrounded by two-dimensional flakes.
{\bf c,} Example scanning electron microscope (SEM) image of the \moss\
nanomaterial, with multiple one-dimensional structures.
{\bf d,} Structure model of a short single-wall \moss\ nanotube segment with
chirality (14,14).
{\bf e,} Example chip lithography drawing, with contact electrodes (gray lines) 
and bond pads (colored squares) designed on top of a nanotube/nanoribbon 
location as detected by optical microscopy (images in black) and position 
markers (light red background pattern).
\label{fig:device}
}
\end{figure*}

Figure \ref{fig:device}(a) shows the raw \moss\ nanomaterial, a mixture of
quasi-one dimensional and quasi-two dimensional structures, as deposited on the
wall of the growth quartz glass ampoule. The material synthesis via vapor phase
deposition is well-established and has been discussed in detail in published
literature \cite{remskar1996, schock2025}. Detail optical and scanning electron
microscope (SEM) images in Figures \ref{fig:device}(b,c) demonstrate the
presence of long and crystalline one-dimensional morphologies, including
multiwall nanotubes, cf. Fig.~\ref{fig:device}(d), and nanoribbons, i.e.,
nanotubes collapsed during growth. The nanomaterial is picked up using a
commercially available adhesive tape and deposited on the surface of a p++ Si /
SiO$_2$ wafer with prefabricated grid markers \cite{novoselov_electric_2004,
schock2023, schock2025}. Subsequently, the surface is imaged in an optical
microscope and the location of nanotubes and nanowires is recorded for contact
electrode design. As contact material, in either case a bilayer bismuth and
gold is deposited, as based on and described in previous work \cite{shen2021a,
schock2023, schock2025}. An additional adhesion layer is not required, since
thermally evaporated bismuth bonds sufficiently to a silicon oxide or hBN
surface. The drawing of Fig.~\ref{fig:device}(e) demonstrates an example of a
resulting chip geometry, with bond pads, leads, and nanotube regions. The
active device region between contacts is chosen to be between
\qtyrange{100}{500}{\nm}.

As already discussed previously \cite{schock2023, schock2025}, while highly
succesful and reproducible for 2d materials \cite{shen2021a}, this technique
leads for nanotubes and nanoribbons to a few good devices but overall a very
large scatter of two-point resistances, with overall insufficient device yield.
Consequently, a number of device fabrication optimizations have been developed 
and tested in combination, listed in the following and discussed in detail 
further below:

\begin{figure*}[tbp]
\begin{center}
\includegraphics{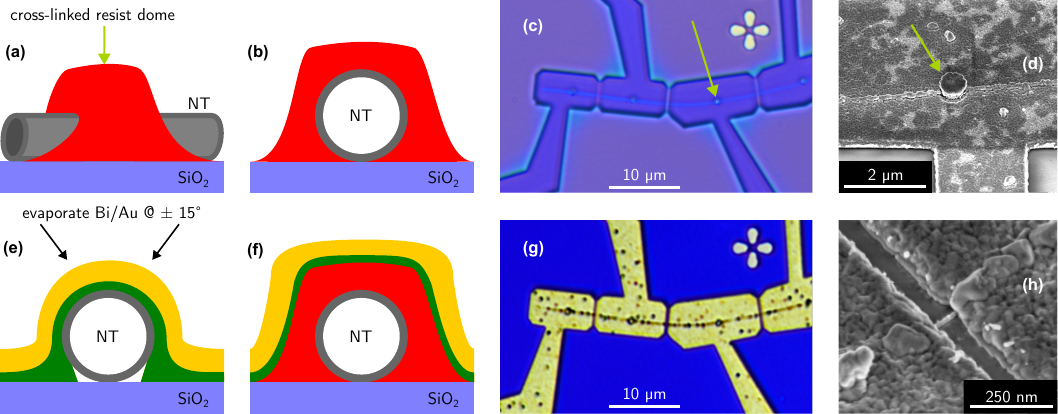}
\end{center}
\caption{{\bf Fabrication details. \quad}
{\bf a, b,}  Sketches illustrating the local crosslinking of PMMA resist to both
attach a nanotube to the surface and provide a smooth surface for contact 
material evaporation.
{\bf c,} Optical microscope image of a \moss\ nanotube device during
fabrication, with one of the cross-linked resist domes indicated by a
green arrow.
{\bf d,} SEM image of a resist dome after evaporation of Bi/Au contacts.
{\bf e, f,} Sketches illustrating the improved coverage of a nanotube by
dual-angle contact material evaporation, on the blank nanotube as well as in
the region of the cross-linked resist dome.
{\bf g,} Optical microscope image of the same device as in {\bf c}, now after
contact material deposition and lift-off.
{\bf h,} SEM image of an active device region, i.e., the measured nanotube 
segment between contact electrodes.
\label{fig:fabrication}
}
\end{figure*}

\begin{itemize}
{\bf \item ``spot anchoring'':} in a small region of the nanotube contact 
area, poly-methyl metacrylate (PMMA) electron beam lithography resist is 
strongly overexposed, cross-linking the resist on top of and around the nanotube 
and creating an effective glue droplet for mechanical stability and possibly a 
smooth surface topology, see Fig.~\ref{fig:device}(a--d)
{\bf \item deposition pressure:} the ultra-high vacuum setup used for the 
bismuth and gold deposition is pumped out overnight before each deposition, 
with the substrate heated to \qty{100}{\degreeCelsius}, to remove adsorbates 
and atmospheric contaminants; to further improve the chamber pressure of the 
evaporation system, titanium is evaporated with a shutter in front of the 
device, resulting in a final pressure of $p\simeq \qty{5e-8}{\milli\bar}$ 
instead of approximately \qty{3e-7}{\milli\bar} without these steps;
{\bf \item substrate heating:} during the contact deposition process, the 
heating of the chip substrate with the \moss\ nanotube to 
\qty{100}{\degreeCelsius} is kept up, potentially enabling recrystallization of 
the bismuth interface layer as discussed below; and finally
{\bf \item tilted-angle deposition:} both materials bismuth and gold are 
evaporated subsequently at two different angles each, i.e. with the substrate 
rotated by $\pm 15^\circ$ from the orthogonal incident setting.
\end{itemize}

Subsequently a lift-off process in hot acetone is performed.

\section*{Results}

\begin{figure*}[tbp]
\begin{center}
\includegraphics{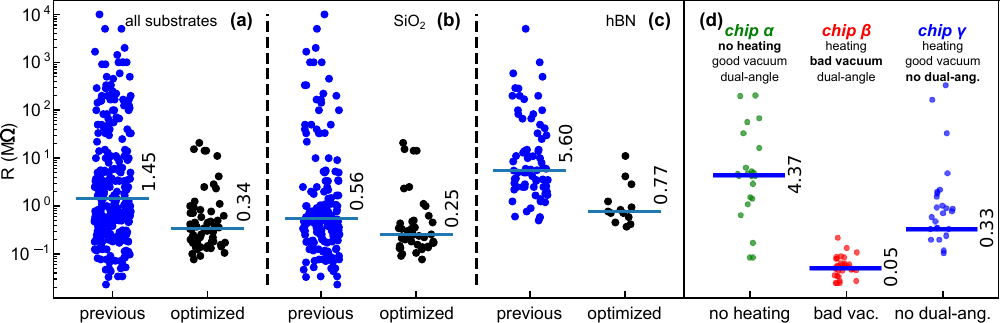}
\end{center}
\caption{{\bf Device resistance distribution
\quad}
{\bf a,} left, blue: distribution of two-point resistances for non-optimized
fabrication, showing a wide scatter, data partially already shown in 
\cite{schock2025}; right, black: optimized fabrication
result using all described steps; the scatter of resistance values is reduced 
by several orders of magnitude. In each case the median value is written out 
and indicated by a horizontal line. 
{\bf b,} analogue plots of the subset of measurements where a 
SiO\textsubscript{2} substrate was directly used; 
{\bf c,} subset of measurements where the nanotube was deposited onto a hBN 
flake. The scatter is reduced similarly in both cases.
{\bf d,} Separate fabrication data set, where in each column one optimization 
step was omitted to obtain insights into the underlying mechanisms; chip 
$\alpha$: no substrate heating, chip $\beta$: bad deposition vacuum (but also 
see text), chip $\gamma$: no dual-angle evaporation.
\label{fig:resistance}
}
\end{figure*}
The two-point resistance characterization at room temperature of the tested
devices is shown in Fig.~\ref{fig:resistance}(a), comparing the straightforward 
contact deposition by evaporation with the procedure optimized using all the 
steps as detailed above. A clear reduction in the scatter of resistance values 
is observed, in conjunction with a decrease in median value. All devices tested 
here with optimized fabrication display an unabiguous source-drain current upon 
application of a bias voltage, with resistances below \qty{30}{\Mohm} and a 
median value of \qty{340}{\kohm}.

Since previous work has focused also on the effect of the substrate on the 
device properties \cite{schock2025}, Figs.~\ref{fig:resistance}(b) and 
\ref{fig:resistance}(c) display a corresponding sorting of the data points. 
Fig.~\ref{fig:resistance}(b) focuses on the case of a bare, amorphous, and
thermally grown silicon oxide, while Fig.~\ref{fig:resistance}(c) focuses on
devices where the nanotube has been transferred onto a hexagonal boron nitride
(hBN) flake \cite{schock2025}. While SiO\textsubscript{2} based devices display
slightly lower resistances, same as in \cite{schock2025}, the effect of the 
optimized contact fabrication is clearly visible in both cases, with 
significantly smaller resistance scatter and a lower median value.

The data of Figure~\ref{fig:resistance}(d) is part of a separate, subsequent 
series of measurements designed to identify the impact of single optimization 
steps on the device properties. Each column corresponds to a single chip with 
several nanotubes and nanoribbons, fabricated with one of the optimizations 
omitted. While the results have to be treated carefully due to fabrication 
irregularities, this still enables us to draw some conclusions. 

Omitting the substrate heating during contact deposition has a clear 
detrimental effect, indicating that this is indeed a crucial detail. The same 
to a slightly lesser degree is true of omitting the dual-angle evaporation. 
A bad vacuum during contact deposition apparently still led to very 
low-resistive devices, see the middle column in Fig.~\ref{fig:resistance}(d); 
unfortunately, large offsets in the data indicate ground loops that were 
unnoticed at the time of measurement. This makes a quantitative analysis here 
unreliable and more measurements would be required for a definite statement; 
still, the overall trend indicates low two-terminal resistances. 

\begin{figure*}[tbp]
\begin{center}
\includegraphics[width=\textwidth]{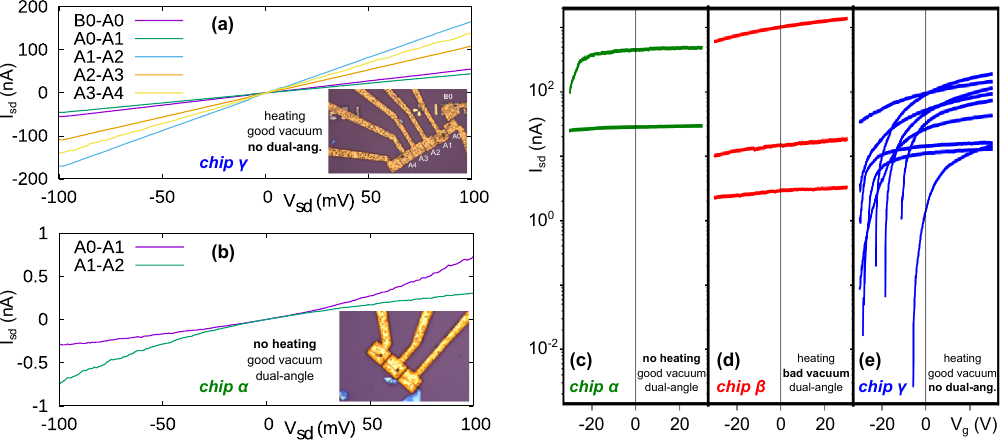}
\end{center}
\caption{{\bf Typical device characteristics
\quad}
{\bf a,} Bias voltage dependence of the current $I(\vsd)$ at $\vg=0$ for 
several nanotube segments on chip $\gamma$ (cf. Fig.~\ref{fig:resistance}(d)). 
Ohmic behaviour is observed. Inset: optical microscope image of the measured 
device. 
{\bf b,} Example bias voltage dependence of the current $I(\vsd)$ at 
$\vg=0$ for two nanotube segments on chip $\alpha$, with clear diode behavior.
Inset: optical microscope image of the measured device.
{\bf c,d,e,} Gate voltage dependence of the device current $I(\vg)$ at 
$\vsd=20\un{mV}$ for selected devices on chips $\alpha$, $\beta$, $\gamma$.
\label{fig:traces}
}
\end{figure*}

Figure~\ref{fig:traces} displays typical further characterization results of 
the chips $\alpha$, $\beta$, and $\gamma$ of the second measurement series. 
All devices on the chips were tested for Ohmic behaviour via recording the bias 
dependence of the current $I(\vsd)$ at $\vg=0$. Predominantly, linear behaviour 
was observed, see, e.g., the measurement of several nanotube segments on chip 
$\gamma$ in Fig.~\ref{fig:traces}(a). The only occurences of clear diode-like 
response were on chip $\alpha$, where the substrate was not heated during 
contact deposition. This is demonstrated in Fig.~\ref{fig:traces}(b) with the 
response of two neighboring segments; the curves nearly coincide when 
mirroring, indicating that the central, shared contact limits the current here.

Fig.~\ref{fig:traces}(c-e) shows the gate voltage response $I(\vg)$ of devices 
on the three chips at fixed $\vsd=20\un{mV}$. While in the recorded range 
$-20\un{V}\le \vg \le 20\un{V}$ a gate voltage dependence consistent with 
electron conduction is visible througout, the response is by far strongest for 
chip $\gamma$ where no dual-angle evaporation was used.

\section*{Discussion}

\subsection*{``Spot anchoring''}

To ensure that the nanotubes adhere to the substrate surface, as well as to
avoid the formation of gaps in the contact material, poly-methyl methacrylate
(PMMA) electron beam resist is spun onto the chip and locally over-exposed by a
factor $\sim 200$. This leads to cross-linking of the PMMA molecules,
effectively turning the positive resist locally into a highly stable
negative-resist and polymer deposition \cite{zailer_crosslinked_1996}, see
Fig.~\ref{fig:fabrication}(b). Fig.~\ref{fig:fabrication}(c) shows an optical 
microscope image of a device with the corresponding droplets (green arrow),
Fig.~\ref{fig:fabrication}(d) is a SEM micrograph of a corresponding contact 
after metal deposition.

The cross-linked PMMA droplets glue the \moss\ nanomaterial to the chip 
surface, significantly improving mechanical stability of the devices. This 
resolves the challenge detailed in \cite{schock2025} (cf. Fig.~8 there), where 
the \moss\ nanostructure frequently tears off the chip during lift-off, taking 
part of the contact metal with it. Furthermore, the polymer deposition can 
assist a smooth surface transition between the chip substrate and the curved 
nanotube surface, see the sketch of Fig.~\ref{fig:fabrication}(b), avoiding 
shadowing effects and gap formation during contact material deposition and 
leading to an unbroken contact film. The precise shape of the result depends 
critically on the resist and contact deposition parameters; as visible in 
Fig.~\ref{fig:fabrication}(d), thicker PMMA leads to mushroom-shaped polymer 
structures.

Chips $\alpha$, $\beta$, $\gamma$ of the second test series each contained 
devices with and without PMMA droplets. Electronic characterization showed no 
clear difference among devices with finite current. This confirms that the main 
impact of the PMMA droplets is to provide mechanical stability and keep the 
nanotube or nanoribbon in place.

\subsection*{Deposition pressure}

The contact material deposition is performed in an UHV chamber system via
electron beam assisted evaporation. In order to reduce contamination with
adsorbates and improve the interface quality, the devices are heated to
$100^\circ\text{C}$ inside the UHV system for $>$ \qty{10}{\hour} before
deposition. Additionally, a short Ti gettering step is performed to further
reduce the pressure of the evaporation system to $p = \qty{5e-8}{\milli\bar}$.

Similar optimization steps were performed in early high-profile works on carbon 
nanotubes that used comparable fabrication \cite{nature-jarillo-herrero-2004, 
nature-jarillo-herrero-2005, private-pablo}. Adsorbates or chemical surface 
modifications can have a strong impact on the interface properties of 
low-dimensional materials and the Schottky barrier \cite{tung_physics_2014}, as 
also attested by a large body of complex and sometimes contradicting published 
experimental results on 2d \moss\ devices. Recent studies have revealed that 
the quality of Ni contacts to \moss\ depends heavily on the vacuum, with 
typically $\sim 3$ times lower contact resistance going from HV to UHV 
deposition conditions \cite{sun_hv_uhv_NiBi_2024}.

The authors of Ref.~\cite{sun_hv_uhv_NiBi_2024} could not identify any 
significant improvement in contact resistance for Bi contacts for different 
vacuum conditions. This is consistent with our observations when skipping the 
vacuum optimization alone (chip $\beta$), where this step turns out not to be 
critical on a smaller test scale, but even led to surprisingly good contacts. 
Further test would be required to confirm or refute whether a good vacuum has 
advantages.

\subsection*{Contact materials}

The choice of a bismuth / gold bilayer as contact material for \moss\ was first
championed in \cite{shen2021a} for flake-based 2d material devices, with
outstanding results. Many earlier attempts at Ohmic contacts to \moss\ had been
unsuccessful due to Fermi level pinning and Schottky barrier formation. Low
work function metals such as titanium or scandium are a possible solution
\cite{das2013, reinhardt2019a}, however tend to react with the \moss\ surface
on deposition \cite{wu2019}, particularly detrimental in the low-dimensional
material case and for low-temperature applications. Doping with copper
\cite{liu_improvement_2018} or yttrium \cite{jiang_yttrium-doping-induced_2024} 
has in general shown good results, but does not make much sense for nanotubes 
due to the small surface area involved and the resulting electrostatic 
potential irregularities. 

Instead, bismuth as semimetal was chosen since its vanishing density of states 
at the Fermi edge avoids the formation of metal-induced gap states and with it 
Fermi level pinning \cite{louie_electronic_1976, monch_valence-band_1998, 
shen2021a}. Calculations indicate that the effective contact resistance results 
from an interplay of the absence of interaction (leading to an absence of 
metal-induced gap states) and the absence of tunneling (leading to a 
suppression of current), close to the optimum for adjacent 2d layers 
\cite{lizzit_ohmic_2023}. For nanotubes, the curvature of the surface as well 
as the resulting more complex contact material deposition and morphology can 
then easily explain the remaining resistance fluctuations.

While antimony delivers even better results for 2d material devices
\cite{li2023}, specifically for nanotubes and nanoribbons measurements have
demonstrated bismuth to be the best choice \cite{schock2025}. A detailed
explanation for this is still missing; one can speculate that a bismuth layer
may be less sensitive to the curved nanotube surface or the precise deposition 
conditions, see also the discussion on recrystallization below.

\subsection*{Deposition protocol}

The contact materials, bismuth and gold, are deposited with \qty{50}{\nm} and
\qty{100}{\nm} at rates of \qty{0.03}{\nm\per\second} and
\qty{0.05}{\nm\per\second}, respectively. The substrate is kept heated at
\qty{100}{\degreeCelsius} and the incidence angle of the deposition is changed
from \qty{-15}{\degree} to \qty{+15}{\degree} after half the layer thickness
has been reached for each material to improve coverage. After deposition, the
device is left inside the UHV chamber until reaching ambient temperature.

Thermal evaporation, while providing high quality thin films, is a directed 
process. As a consequence, shadowing and capping effects can occur when surface 
features such as the nanotubes here are of similar or larger scale as the film 
thickness \cite{schock2025}, an effect that is in literature also deliberately 
used for device structuring \cite{chen_gate-tunable_2023}. Dual angle 
evaporation reduces this, with the material reaching into gaps and below 
overhangs. Its specific advantage is confirmed by the measurement of chip 
$\gamma$, where without dual angle evaporation clearly increased resistance 
scatter and median value is observed, see Fig.~\ref{fig:resistance}(d). 
Additionally, without dual-angle evaporation, a stronger gate effect is 
visible, see Fig.~\ref{fig:traces}(e). As unfortunate side effect, better 
coverage of the nanotube or nanoribbon appears to lead to partial shielding of 
the gate voltage.

Bismuth has comparatively low melting point of $T=271^\circ\text{C}$. It tends 
to form crystallites during thin film deposition \cite{kumari_effects_2007}; 
this is visible in, e.g., Fig.~\ref{fig:fabrication}(d,h) both for the contact 
surface away from the nanotube and for additional nucleation directly on top of 
it, even with the additional gold capping layer. Early research has already 
addressed the dependence of the crystal formation on the substrate temperature 
\cite{partin_growth_1989, kumari_effects_2007}, indicating that higher 
temperature leads to better film quality. This is consistent with the 
observation that chip $\alpha$ of the second test series, where substrate 
heating was omitted, was the only one to display Schottky diode behaviour, see 
Fig.~\ref{fig:traces}(b). Furthermore, for our bismuth / gold bilayer, it is 
likely that heating effects during the subsequent gold deposition lead to 
(partial) melting and recrystallization, improving contact transparency 
\cite{private-ljl}. The formation of the gold/bismuth intermetallic compound 
maldonite (Au\textsubscript{2}Bi) at the interface \cite{okamoto_aubi_1983} is 
also possible and has likely been observed by us in a single device where the 
contacts surprisingly displayed superconductivity phenomena at cryogenic 
temperature \cite{dehaas1931super}, but seems to be the exception.

\subsection*{Substrate}

A surprising observation, here as already in \cite{schock2025}, is the 
statistically clearly higher device resistance in case of a hexagonal boron 
nitride (hBN) substrate. While the amorphous \sioo\ is known for its dangling 
bonds and surface charges, hBN has been the material of choice to precisely 
avoid such effects. As a possible cause, the more complex fabrication of 
hBN-based devices stands out: here, the hBN flakes have to be exfoliated and 
deposited onto the chip substrate, and subsequently also the \moss\ 
nanomaterial has to be transferred again \cite{schock2025}. Each such step 
increases the risk of surface contamination and structural damage.

\section*{Conclusions and Outlook}

We have managed to improve the fabrication of metallic contacts to \moss\ 
nanotubes and nanoribbons such that we consistently achieve Ohmic contacts, 
with a median 2-point resistance of a segment of $340\un{k}\Omega$ in a large 
fabrication series. The impact of separate improvements is tested and 
discussed. Decisive steps include local PMMA crosslinking for mechanical 
stabilization, substrate heating to $100^\circ\text{C}$ during contact 
deposition, and dual-angle evaporation to improve coverage of the non-planar 
nanomaterial. The precise impact of deposition pressure is so far unclear. 

In previous work, the integration of \moss\ nanotubes and nanoribbons into 
complex devices was challenging due to low fabrication yield. With reliable 
contact formation, the path towards 1d \moss\ transistors as well as 1d \moss\ 
quantum electronics is now open.

What has not been adressed here is the intrinsic surface quality of the \moss\
nanotubes and nanoribbons as well as the impact of their geometry. Sulfur 
vacancies have been reported to be beneficial for n-type contacts 
\cite{mcdonnell_svacancy_ndoping_2014}, but also modify the nanomaterial 
properties in between them; fabrication in inert environment and 
treatment with H\textsubscript{2}S \cite{peto_spontaneous_2018} or thioles 
\cite{schwarz_thiol-based_2023} may help here. Wall curvature, inner 
substructures, and deformation on substrate deposition modify the electronic 
properties \cite{malok_edge_2025, malok_electrical_2025}. A smaller number of 
nanotube walls will likely improve reproducibility of device properties; free 
single-wall \moss\ nanotubes are thermodynamically unstable 
\cite{khadiev_misfit_2024}, but the recent development of coaxial 
heterostructures provides a path to solve this problem 
\cite{xiang_one-dimensional_2020}.

In the end, while challenges in nanomaterial chemistry and device technology 
remain, the appeal of truly one-dimensional \moss\ for both traditional and 
quantum electronics remains very much undiminished, and a significant step has 
been made towards device implementation.

\section*{Acknowledgments}
The authors gratefully acknowledge funding by the DFG via grants Hu 1808/4-1
(project id 438638106) and Hu 1808/6-1 (project id 438640730) and by the
Slovenian Research Agency via grants P1-0099 and PR-11224. M.~K. was funded by 
the DFG via CRC/SFB 1277 (project id 314695032). We would like to thank 
Lain-Jong Li for insightful discussions and Ch.~Strunk, D.~Weiss, and 
F.~Kuemmeth for the use of experimental facilities. The measurement data was 
recorded using Lab::Measurement \cite{labmeasurement}.

\section*{Author contributions}
The \moss\ nanomaterial was grown at Jozef Stefan Institute by M.~M. The
devices were fabricated by R.~T.~K.~S., S.~O., and K.~F. with help from M.~K.
The room temperature current measurements were performed by R.~T.~K.~S., S.~O.,
and K.~F. The manuscript was written by A.~K.~H. and R.~T.~K.~S. and reviewed
and revised by all coauthors. The project was supervised by M.~R. in Ljubljana
and by A.~K.~H. in Regensburg.

\section*{Competing interests}
The authors declare no competing interests.

\vspace*{0.7cm}
\section*{Data availability}
The datasets generated during and/or analysed during this study are available
from the corresponding author on reasonable request.

\bibliographystyle{naturemag}
\bibliography{paper}

\begin{thebibliography}{10}
\expandafter\ifx\csname url\endcsname\relax
  \def\url#1{\texttt{#1}}\fi
\expandafter\ifx\csname urlprefix\endcsname\relax\def\urlprefix{URL }\fi
\providecommand{\bibinfo}[2]{#2}
\providecommand{\eprint}[2][]{\url{#2}}

\bibitem{novoselov_electric_2004}
\bibinfo{author}{Novoselov, K.~S.} \emph{et~al.}
\newblock \bibinfo{title}{Electric field effect in atomically thin carbon
  films}.
\newblock \emph{\bibinfo{journal}{Science}} \textbf{\bibinfo{volume}{306}},
  \bibinfo{pages}{666--669} (\bibinfo{year}{2004}).
\newblock \urlprefix\url{https://www.science.org/doi/10.1126/science.1102896}.

\bibitem{novoselov_two-dimensional_2005}
\bibinfo{author}{Novoselov, K.~S.} \emph{et~al.}
\newblock \bibinfo{title}{Two-dimensional atomic crystals}.
\newblock \emph{\bibinfo{journal}{Proceedings of the National Academy of
  Sciences}} \textbf{\bibinfo{volume}{102}}, \bibinfo{pages}{10451--10453}
  (\bibinfo{year}{2005}).
\newblock \urlprefix\url{https://www.pnas.org/content/102/30/10451}.

\bibitem{das_transistors_2021}
\bibinfo{author}{Das, S.} \emph{et~al.}
\newblock \bibinfo{title}{Transistors based on two-dimensional materials for
  future integrated circuits}.
\newblock \emph{\bibinfo{journal}{Nature Electronics}}
  \textbf{\bibinfo{volume}{4}}, \bibinfo{pages}{786--799}
  (\bibinfo{year}{2021}).
\newblock \urlprefix\url{https://www.nature.com/articles/s41928-021-00670-1}.

\bibitem{dorow_exploring_2023}
\bibinfo{author}{Dorow, C.~J.} \emph{et~al.}
\newblock \bibinfo{title}{Exploring manufacturability of novel {2D} channel
  materials: 300mm wafer-scale {2D} {NMOS} \& {PMOS} using
  {MoS\textsubscript{2}}, {WS\textsubscript{2}}, \& {WSe\textsubscript{2}}}.
\newblock In \emph{\bibinfo{booktitle}{2023 {International} {Electron}
  {Devices} {Meeting} ({IEDM})}}, \bibinfo{pages}{1--4} (\bibinfo{year}{2023}).
\newblock \urlprefix\url{https://dx.doi.org/10.1109/IEDM45741.2023.10413874}.

\bibitem{chou_status_2023}
\bibinfo{author}{Chou, A.-S.} \emph{et~al.}
\newblock \bibinfo{title}{Status and performance of integration modules toward
  scaled {CMOS} with transition metal dichalcogenide channel}.
\newblock In \emph{\bibinfo{booktitle}{2023 {International} {Electron}
  {Devices} {Meeting} ({IEDM})}}, \bibinfo{pages}{1--4} (\bibinfo{year}{2023}).
\newblock \urlprefix\url{https://dx.doi.org/10.1109/IEDM45741.2023.10413779}.

\bibitem{jing_gate-controlled_2022}
\bibinfo{author}{Jing, F.-M.} \emph{et~al.}
\newblock \bibinfo{title}{Gate-controlled quantum dots based on {2D}
  materials}.
\newblock \emph{\bibinfo{journal}{Advanced Quantum Technologies}}
  \textbf{\bibinfo{volume}{5}}, \bibinfo{pages}{2100162}
  (\bibinfo{year}{2022}).
\newblock \urlprefix\url{https://dx.doi.org/10.1002/qute.202100162}.

\bibitem{liu_2d_2019}
\bibinfo{author}{Liu, X.} \& \bibinfo{author}{Hersam, M.~C.}
\newblock \bibinfo{title}{{2D} materials for quantum information science}.
\newblock \emph{\bibinfo{journal}{Nature Reviews Materials}}
  \textbf{\bibinfo{volume}{4}}, \bibinfo{pages}{669--684}
  (\bibinfo{year}{2019}).
\newblock \urlprefix\url{https://www.nature.com/articles/s41578-019-0136-x}.

\bibitem{qiu_recent_2021}
\bibinfo{author}{Qiu, D.} \emph{et~al.}
\newblock \bibinfo{title}{Recent advances in {2D} superconductors}.
\newblock \emph{\bibinfo{journal}{Advanced Materials}}
  \textbf{\bibinfo{volume}{33}}, \bibinfo{pages}{2006124}
  (\bibinfo{year}{2021}).
\newblock
  \urlprefix\url{https://onlinelibrary.wiley.com/doi/abs/10.1002/adma.202006124}.

\bibitem{hills_modern_2019}
\bibinfo{author}{Hills, G.} \emph{et~al.}
\newblock \bibinfo{title}{Modern microprocessor built from complementary carbon
  nanotube transistors}.
\newblock \emph{\bibinfo{journal}{Nature}} \textbf{\bibinfo{volume}{572}},
  \bibinfo{pages}{595--602} (\bibinfo{year}{2019}).
\newblock \urlprefix\url{https://www.nature.com/articles/s41586-019-1493-8}.

\bibitem{si_carbon-nanotube-based_2024}
\bibinfo{author}{Si, J.} \emph{et~al.}
\newblock \bibinfo{title}{A carbon-nanotube-based tensor processing unit}.
\newblock \emph{\bibinfo{journal}{Nature Electronics}}
  \textbf{\bibinfo{volume}{7}}, \bibinfo{pages}{684--693}
  (\bibinfo{year}{2024}).
\newblock \urlprefix\url{https://www.nature.com/articles/s41928-024-01211-2}.

\bibitem{science-stern-2013}
\bibinfo{author}{Stern, A.} \& \bibinfo{author}{Lindner, N.~H.}
\newblock \bibinfo{title}{Topological quantum computation—from basic concepts
  to first experiments}.
\newblock \emph{\bibinfo{journal}{Science}} \textbf{\bibinfo{volume}{339}},
  \bibinfo{pages}{1179--1184} (\bibinfo{year}{2013}).
\newblock \urlprefix\url{https://www.science.org/doi/10.1126/science.1231473}.

\bibitem{science-frolov-2026}
\bibinfo{author}{Frolov, S.~M.} \emph{et~al.}
\newblock \bibinfo{title}{Data sharing helps avoid “smoking gun” claims of
  topological milestones}.
\newblock \emph{\bibinfo{journal}{Science}} \textbf{\bibinfo{volume}{391}},
  \bibinfo{pages}{137--142} (\bibinfo{year}{2026}).
\newblock
  \urlprefix\url{https://www.science.org/doi/abs/10.1126/science.adk9181}.

\bibitem{xiang_one-dimensional_2020}
\bibinfo{author}{Xiang, R.} \emph{et~al.}
\newblock \bibinfo{title}{One-dimensional van der {Waals} heterostructures}.
\newblock \emph{\bibinfo{journal}{Science}} \textbf{\bibinfo{volume}{367}},
  \bibinfo{pages}{537--542} (\bibinfo{year}{2020}).
\newblock \urlprefix\url{https://www.science.org/doi/10.1126/science.aaz2570}.

\bibitem{remskar1996}
\bibinfo{author}{Remskar, M.}, \bibinfo{author}{Skraba, Z.},
  \bibinfo{author}{Cl{\'e}ton, F.}, \bibinfo{author}{Sanjin{\'e}s, R.} \&
  \bibinfo{author}{L{\'e}vy, F.}
\newblock \bibinfo{title}{{MoS\textsubscript{2}} as microtubes}.
\newblock \emph{\bibinfo{journal}{Appl. Phys. Lett.}}
  \textbf{\bibinfo{volume}{69}}, \bibinfo{pages}{351--353}
  (\bibinfo{year}{1996}).
\newblock \urlprefix\url{https://dx.doi.org/10.1063/1.118057}.

\bibitem{kazanov_multiwall_2018}
\bibinfo{author}{Kazanov, D.~R.} \emph{et~al.}
\newblock \bibinfo{title}{Multiwall {MoS\textsubscript{2}} tubes as optical
  resonators}.
\newblock \emph{\bibinfo{journal}{Applied Physics Letters}}
  \textbf{\bibinfo{volume}{113}}, \bibinfo{pages}{101106}
  (\bibinfo{year}{2018}).
\newblock \urlprefix\url{https://dx.doi.org/10.1063/1.5047792}.

\bibitem{remskar2022}
\bibinfo{author}{Remskar, M.} \emph{et~al.}
\newblock \bibinfo{title}{Confinement related phenomena in
  {MoS\textsubscript{2}} tubular structures grown from vapour phase}.
\newblock \emph{\bibinfo{journal}{Israel Journal of Chemistry}}
  \textbf{\bibinfo{volume}{62}}, \bibinfo{pages}{e202100100}
  (\bibinfo{year}{2022}).
\newblock \urlprefix\url{https://dx.doi.org/10.1002/ijch.202100100}.

\bibitem{fathipour2015}
\bibinfo{author}{Fathipour, S.} \emph{et~al.}
\newblock \bibinfo{title}{Synthesized multiwall {MoS\textsubscript{2}} nanotube
  and nanoribbon field-effect transistors}.
\newblock \emph{\bibinfo{journal}{Appl. Phys. Lett.}}
  \textbf{\bibinfo{volume}{106}}, \bibinfo{pages}{022114}
  (\bibinfo{year}{2015}).
\newblock \urlprefix\url{https://dx.doi.org/10.1063/1.4906066}.

\bibitem{zphys-schottky-1939}
\bibinfo{author}{Schottky, W.}
\newblock \bibinfo{title}{{Zur Halbleitertheorie der Sperrschicht- und
  Spitzengleichrichter}}.
\newblock \emph{\bibinfo{journal}{Zeitschrift für Physik}}
  \textbf{\bibinfo{volume}{113}}, \bibinfo{pages}{367--414}
  (\bibinfo{year}{1939}).
\newblock \urlprefix\url{https://dx.doi.org/10.1007/BF01340116}.

\bibitem{tung_physics_2014}
\bibinfo{author}{Tung, R.~T.}
\newblock \bibinfo{title}{The physics and chemistry of the {Schottky} barrier
  height}.
\newblock \emph{\bibinfo{journal}{Applied Physics Reviews}}
  \textbf{\bibinfo{volume}{1}}, \bibinfo{pages}{011304} (\bibinfo{year}{2014}).
\newblock \urlprefix\url{https://aip.scitation.org/doi/10.1063/1.4858400}.

\bibitem{louie_electronic_1976}
\bibinfo{author}{Louie, S.~G.} \& \bibinfo{author}{Cohen, M.~L.}
\newblock \bibinfo{title}{Electronic structure of a metal-semiconductor
  interface}.
\newblock \emph{\bibinfo{journal}{Physical Review B}}
  \textbf{\bibinfo{volume}{13}}, \bibinfo{pages}{2461--2469}
  (\bibinfo{year}{1976}).
\newblock \urlprefix\url{https://dx.doi.org/10.1103/PhysRevB.13.2461}.

\bibitem{monch_valence-band_1998}
\bibinfo{author}{Mönch, W.}
\newblock \bibinfo{title}{Valence-band offsets and {Schottky} barrier heights
  of layered semiconductors explained by interface-induced gap states}.
\newblock \emph{\bibinfo{journal}{Applied Physics Letters}}
  \textbf{\bibinfo{volume}{72}}, \bibinfo{pages}{1899--1901}
  (\bibinfo{year}{1998}).
\newblock \urlprefix\url{https://dx.doi.org/10.1063/1.121220}.

\bibitem{nl-gong-2014}
\bibinfo{author}{Gong, C.}, \bibinfo{author}{Colombo, L.},
  \bibinfo{author}{Wallace, R.~M.} \& \bibinfo{author}{Cho, K.}
\newblock \bibinfo{title}{The unusual mechanism of partial {Fermi} level
  pinning at metal–{MoS\textsubscript{2}} interfaces}.
\newblock \emph{\bibinfo{journal}{Nano Letters}} \textbf{\bibinfo{volume}{14}},
  \bibinfo{pages}{1714--1720} (\bibinfo{year}{2014}).
\newblock \urlprefix\url{https://dx.doi.org/10.1021/nl403465v}.

\bibitem{kim_fermi_2017}
\bibinfo{author}{Kim, C.} \emph{et~al.}
\newblock \bibinfo{title}{{Fermi} level pinning at electrical metal contacts of
  monolayer molybdenum dichalcogenides}.
\newblock \emph{\bibinfo{journal}{ACS Nano}} \textbf{\bibinfo{volume}{11}},
  \bibinfo{pages}{1588--1596} (\bibinfo{year}{2017}).
\newblock \urlprefix\url{https://dx.doi.org/10.1021/acsnano.6b07159}.

\bibitem{jpcc-sotthewes-2019}
\bibinfo{author}{Sotthewes, K.} \emph{et~al.}
\newblock \bibinfo{title}{Universal {Fermi}-level pinning in transition-metal
  dichalcogenides}.
\newblock \emph{\bibinfo{journal}{The Journal of Physical Chemistry C}}
  \textbf{\bibinfo{volume}{123}}, \bibinfo{pages}{5411--5420}
  (\bibinfo{year}{2019}).
\newblock \urlprefix\url{https://dx.doi.org/10.1021/acs.jpcc.8b10971}.

\bibitem{shen2021a}
\bibinfo{author}{Shen, P.-C.} \emph{et~al.}
\newblock \bibinfo{title}{Ultralow contact resistance between semimetal and
  monolayer semiconductors}.
\newblock \emph{\bibinfo{journal}{Nature}} \textbf{\bibinfo{volume}{593}},
  \bibinfo{pages}{211--217} (\bibinfo{year}{2021}).
\newblock \urlprefix\url{https://dx.doi.org/10.1038/s41586-021-03472-9}.

\bibitem{schock2023}
\bibinfo{author}{Schock, R. T.~K.} \emph{et~al.}
\newblock \bibinfo{title}{Non-destructive low-temperature contacts to
  {MoS\textsubscript{2}} nanoribbon and nanotube quantum dots}.
\newblock \emph{\bibinfo{journal}{Advanced Materials}}
  \textbf{\bibinfo{volume}{35}}, \bibinfo{pages}{2209333}
  (\bibinfo{year}{2023}).
\newblock \urlprefix\url{https://dx.doi.org/10.1002/adma.202209333}.

\bibitem{li2023}
\bibinfo{author}{Li, W.} \emph{et~al.}
\newblock \bibinfo{title}{Approaching the quantum limit in two-dimensional
  semiconductor contacts}.
\newblock \emph{\bibinfo{journal}{Nature}} \textbf{\bibinfo{volume}{613}},
  \bibinfo{pages}{274--279} (\bibinfo{year}{2023}).
\newblock \urlprefix\url{https://dx.doi.org/10.1038/s41586-022-05431-4}.

\bibitem{schock2025}
\bibinfo{author}{Schock, R. T.~K.} \emph{et~al.}
\newblock \bibinfo{title}{Material transfer and contact optimization in
  {MoS\textsubscript{2}} nanotube devices}.
\newblock \emph{\bibinfo{journal}{physica status solidi (b)}}
  \textbf{\bibinfo{volume}{262}}, \bibinfo{pages}{2400366}
  (\bibinfo{year}{2025}).
\newblock
  \urlprefix\url{https://onlinelibrary.wiley.com/doi/abs/10.1002/pssb.202400366}.

\bibitem{zailer_crosslinked_1996}
\bibinfo{author}{Zailer, I.}, \bibinfo{author}{Frost, J. E.~F.},
  \bibinfo{author}{Chabasseur-Molyneux, V.}, \bibinfo{author}{Ford, C. J.~B.}
  \& \bibinfo{author}{Pepper, M.}
\newblock \bibinfo{title}{Crosslinked {PMMA} as a high-resolution negative
  resist for electron beam lithography and applications for physics of
  low-dimensional structures}.
\newblock \emph{\bibinfo{journal}{Semiconductor Science and Technology}}
  \textbf{\bibinfo{volume}{11}}, \bibinfo{pages}{1235} (\bibinfo{year}{1996}).
\newblock \urlprefix\url{https://dx.doi.org/10.1088/0268-1242/11/8/021}.

\bibitem{nature-jarillo-herrero-2004}
\bibinfo{author}{Jarillo-Herrero, P.}, \bibinfo{author}{Sapmaz, S.},
  \bibinfo{author}{Dekker, C.}, \bibinfo{author}{Kouwenhoven, L.~P.} \&
  \bibinfo{author}{van~der Zant, H. S.~J.}
\newblock \bibinfo{title}{Electron-hole symmetry in a semiconducting carbon
  nanotube quantum dot}.
\newblock \emph{\bibinfo{journal}{Nature}} \textbf{\bibinfo{volume}{429}},
  \bibinfo{pages}{389--392} (\bibinfo{year}{2004}).
\newblock \urlprefix\url{https://www.nature.com/articles/nature02568}.

\bibitem{nature-jarillo-herrero-2005}
\bibinfo{author}{Jarillo-Herrero, P.} \emph{et~al.}
\newblock \bibinfo{title}{Orbital {Kondo} effect in carbon nanotubes}.
\newblock \emph{\bibinfo{journal}{Nature}} \textbf{\bibinfo{volume}{434}},
  \bibinfo{pages}{484--488} (\bibinfo{year}{2005}).
\newblock \urlprefix\url{https://www.nature.com/articles/nature03422}.

\bibitem{private-pablo}
\bibinfo{author}{Jarillo-Herrero, P.} (\bibinfo{year}{2007}).
\newblock \bibinfo{note}{Private communication}.

\bibitem{sun_hv_uhv_NiBi_2024}
\bibinfo{author}{Sun, Z.} \emph{et~al.}
\newblock \bibinfo{title}{Low contact resistance on monolayer
  {MoS\textsubscript{2}} field-effect transistors achieved by {CMOS}-compatible
  metal contacts}.
\newblock \emph{\bibinfo{journal}{ACS Nano}} \textbf{\bibinfo{volume}{18}},
  \bibinfo{pages}{22444–22453} (\bibinfo{year}{2024}).
\newblock \urlprefix\url{https://dx.doi.org/10.1021/acsnano.4c07267}.

\bibitem{das2013}
\bibinfo{author}{Das, S.}, \bibinfo{author}{Chen, H.-Y.},
  \bibinfo{author}{Penumatcha, A.~V.} \& \bibinfo{author}{Appenzeller, J.}
\newblock \bibinfo{title}{High performance multilayer {MoS\textsubscript{2}}
  transistors with scandium contacts}.
\newblock \emph{\bibinfo{journal}{Nano Letters}} \textbf{\bibinfo{volume}{13}},
  \bibinfo{pages}{100--105} (\bibinfo{year}{2013}).
\newblock \urlprefix\url{https://dx.doi.org/10.1021/nl303583v}.

\bibitem{reinhardt2019a}
\bibinfo{author}{Reinhardt, S.}, \bibinfo{author}{Pirker, L.},
  \bibinfo{author}{B{\"a}uml, C.}, \bibinfo{author}{Rem{\v s}kar, M.} \&
  \bibinfo{author}{H{\"u}ttel, A.~K.}
\newblock \bibinfo{title}{Coulomb blockade spectroscopy of a
  {MoS\textsubscript{2}} nanotube}.
\newblock \emph{\bibinfo{journal}{physica status solidi (RRL) -- Rapid Research
  Letters}} \textbf{\bibinfo{volume}{13}}, \bibinfo{pages}{1900251}
  (\bibinfo{year}{2019}).
\newblock \urlprefix\url{https://dx.doi.org/10.1002/pssr.201900251}.

\bibitem{wu2019}
\bibinfo{author}{Wu, R.~J.} \emph{et~al.}
\newblock \bibinfo{title}{Visualizing the metal-{MoS\textsubscript{2}} contacts
  in two-dimensional field-effect transistors with atomic resolution}.
\newblock \emph{\bibinfo{journal}{Phys. Rev. Materials}}
  \textbf{\bibinfo{volume}{3}}, \bibinfo{pages}{111001} (\bibinfo{year}{2019}).
\newblock \urlprefix\url{https://dx.doi.org/10.1103/PhysRevMaterials.3.111001}.

\bibitem{liu_improvement_2018}
\bibinfo{author}{Liu, M.} \emph{et~al.}
\newblock \bibinfo{title}{Improvement of metal-semiconductor contact from
  {Schottky} to {Ohmic} by {Cu} doping in transition metal dichalcogenide
  transistors}.
\newblock In \emph{\bibinfo{booktitle}{2018 {IEEE} 13th {Nanotechnology}
  {Materials} and {Devices} {Conference}}}, \bibinfo{pages}{1--4}
  (\bibinfo{publisher}{IEEE}, \bibinfo{address}{Portland, OR, USA},
  \bibinfo{year}{2018}).
\newblock \urlprefix\url{https://ieeexplore.ieee.org/document/8605891/}.

\bibitem{jiang_yttrium-doping-induced_2024}
\bibinfo{author}{Jiang, J.} \emph{et~al.}
\newblock \bibinfo{title}{Yttrium-doping-induced metallization of molybdenum
  disulfide for {Ohmic} contacts in two-dimensional transistors}.
\newblock \emph{\bibinfo{journal}{Nature Electronics}}
  \textbf{\bibinfo{volume}{7}}, \bibinfo{pages}{545--556}
  (\bibinfo{year}{2024}).
\newblock \urlprefix\url{https://www.nature.com/articles/s41928-024-01176-2}.

\bibitem{lizzit_ohmic_2023}
\bibinfo{author}{Lizzit, D.}, \bibinfo{author}{Khakbaz, P.},
  \bibinfo{author}{Driussi, F.}, \bibinfo{author}{Pala, M.} \&
  \bibinfo{author}{Esseni, D.}
\newblock \bibinfo{title}{{Ohmic} behavior in metal contacts to n/p-type
  transition-metal dichalcogenides: {Schottky} versus tunneling barrier
  trade-off}.
\newblock \emph{\bibinfo{journal}{ACS Applied Nano Materials}}
  (\bibinfo{year}{2023}).
\newblock \urlprefix\url{https://doi.org/10.1021/acsanm.3c00166}.

\bibitem{chen_gate-tunable_2023}
\bibinfo{author}{Chen, Y.} \emph{et~al.}
\newblock \bibinfo{title}{Gate-tunable superconductivity in hybrid
  {InSb}–{Pb} nanowires}.
\newblock \emph{\bibinfo{journal}{Applied Physics Letters}}
  \textbf{\bibinfo{volume}{123}}, \bibinfo{pages}{082601}
  (\bibinfo{year}{2023}).
\newblock \urlprefix\url{https://dx.doi.org/10.1063/5.0155663}.

\bibitem{kumari_effects_2007}
\bibinfo{author}{Kumari, L.} \emph{et~al.}
\newblock \bibinfo{title}{Effects of deposition temperature and thickness on
  the structural properties of thermal evaporated bismuth thin films}.
\newblock \emph{\bibinfo{journal}{Applied Surface Science}}
  \textbf{\bibinfo{volume}{253}}, \bibinfo{pages}{5931--5938}
  (\bibinfo{year}{2007}).
\newblock \urlprefix\url{https://dx.doi.org/10.1016/j.apsusc.2006.12.125}.

\bibitem{partin_growth_1989}
\bibinfo{author}{Partin, D.~L.}, \bibinfo{author}{Thrush, C.~M.},
  \bibinfo{author}{Heremans, J.}, \bibinfo{author}{Morelli, D.~T.} \&
  \bibinfo{author}{Olk, C.~H.}
\newblock \bibinfo{title}{Growth and characterization of epitaxial bismuth
  films}.
\newblock \emph{\bibinfo{journal}{Journal of Vacuum Science \& Technology B:
  Microelectronics Processing and Phenomena}} \textbf{\bibinfo{volume}{7}},
  \bibinfo{pages}{348--353} (\bibinfo{year}{1989}).
\newblock \urlprefix\url{https://doi.org/10.1116/1.584748}.

\bibitem{private-ljl}
\bibinfo{author}{Li, L.-J.} (\bibinfo{year}{2024}).
\newblock \bibinfo{note}{Private communication}.

\bibitem{okamoto_aubi_1983}
\bibinfo{author}{Okamoto, H.} \& \bibinfo{author}{Massalski, T.~B.}
\newblock \bibinfo{title}{The {Au}-{Bi} (gold-bismuth) system}.
\newblock \emph{\bibinfo{journal}{Bulletin of Alloy Phase Diagrams}}
  \textbf{\bibinfo{volume}{4}}, \bibinfo{pages}{401--407}
  (\bibinfo{year}{1983}).
\newblock \urlprefix\url{https://dx.doi.org/10.1007/BF02868093}.

\bibitem{dehaas1931super}
\bibinfo{author}{de~Haas, W.~J.} \& \bibinfo{author}{Jurriaanse, F.}
\newblock \bibinfo{title}{The super-conductivity of gold-bismuth}.
\newblock \emph{\bibinfo{journal}{Naturwissenschaften}}
  \textbf{\bibinfo{volume}{19}}, \bibinfo{pages}{706--706}
  (\bibinfo{year}{1931}).

\bibitem{mcdonnell_svacancy_ndoping_2014}
\bibinfo{author}{McDonnell, S.}, \bibinfo{author}{Addou, R.},
  \bibinfo{author}{Buie, C.}, \bibinfo{author}{Wallace, R.~M.} \&
  \bibinfo{author}{Hinkle, C.~L.}
\newblock \bibinfo{title}{Defect-dominated doping and contact resistance in
  {MoS\textsubscript{2}}}.
\newblock \emph{\bibinfo{journal}{ACS Nano}} \textbf{\bibinfo{volume}{8}},
  \bibinfo{pages}{2880--2888} (\bibinfo{year}{2014}).
\newblock \urlprefix\url{https://dx.doi.org/10.1021/nn500044q}.

\bibitem{peto_spontaneous_2018}
\bibinfo{author}{Pető, J.} \emph{et~al.}
\newblock \bibinfo{title}{Spontaneous doping of the basal plane of
  {MoS\textsubscript{2}} single layers through oxygen substitution under
  ambient conditions}.
\newblock \emph{\bibinfo{journal}{Nature Chemistry}}
  \textbf{\bibinfo{volume}{10}}, \bibinfo{pages}{1246--1251}
  (\bibinfo{year}{2018}).
\newblock \urlprefix\url{https://www.nature.com/articles/s41557-018-0136-2}.

\bibitem{schwarz_thiol-based_2023}
\bibinfo{author}{Schwarz, A.} \emph{et~al.}
\newblock \bibinfo{title}{Thiol-based defect healing of {WSe\textsubscript{2}}
  and {WS\textsubscript{2}}}.
\newblock \emph{\bibinfo{journal}{npj 2D Materials and Applications}}
  \textbf{\bibinfo{volume}{7}}, \bibinfo{pages}{1--9} (\bibinfo{year}{2023}).
\newblock \urlprefix\url{https://www.nature.com/articles/s41699-023-00421-0}.

\bibitem{malok_edge_2025}
\bibinfo{author}{Malok, M.}, \bibinfo{author}{Jelenc, J.},
  \bibinfo{author}{Krajnc, A.~P.} \& \bibinfo{author}{Remškar, M.}
\newblock \bibinfo{title}{Edge and defect effects on charge distribution in
  collapsed {MoS\textsubscript{2}} nanotubes}.
\newblock \emph{\bibinfo{journal}{Nanoscale Advances}}
  \textbf{\bibinfo{volume}{7}}, \bibinfo{pages}{8161--8169}
  (\bibinfo{year}{2025}).
\newblock \urlprefix\url{https://dx.doi.org/10.1039/D5NA00771B}.

\bibitem{malok_electrical_2025}
\bibinfo{author}{Malok, M.}, \bibinfo{author}{Jelenc, J.} \&
  \bibinfo{author}{Remškar, M.}
\newblock \bibinfo{title}{Electrical properties of collapsed
  {MoS\textsubscript{2}} nanotubes}.
\newblock \emph{\bibinfo{journal}{Nanoscale}} \textbf{\bibinfo{volume}{17}},
  \bibinfo{pages}{12361--12370} (\bibinfo{year}{2025}).
\newblock \urlprefix\url{https://dx.doi.org/10.1039/D5NR00284B}.

\bibitem{khadiev_misfit_2024}
\bibinfo{author}{Khadiev, A.} \emph{et~al.}
\newblock \bibinfo{title}{Misfit layered compounds: Insights into chemical,
  kinetic, and thermodynamic stability of nanophases}.
\newblock \emph{\bibinfo{journal}{Accounts of Chemical Research}}
  \textbf{\bibinfo{volume}{57}}, \bibinfo{pages}{3243--3253}
  (\bibinfo{year}{2024}).
\newblock
  \urlprefix\url{https://pubs.acs.org/doi/10.1021/acs.accounts.4c00412}.

\bibitem{labmeasurement}
\bibinfo{author}{Reinhardt, S.} \emph{et~al.}
\newblock \bibinfo{title}{{Lab::Measurement} --- a portable and extensible
  framework for controlling lab equipment and conducting measurements}.
\newblock \emph{\bibinfo{journal}{Computer Physics Communications}}
  \textbf{\bibinfo{volume}{234}}, \bibinfo{pages}{216} (\bibinfo{year}{2019}).
\newblock \urlprefix\url{https://dx.doi.org/10.1016/j.cpc.2018.07.024}.

\end{thebibliography}

\end{document}